\documentclass[aps,pre,twocolumn,groupedaddress,showpacs]{revtex4}
\usepackage{graphicx}
\usepackage{amssymb}

\begin{document}

\title{Statistical characterization of the forces on spheres in an
upflow of air}


\author{R.P. Ojha, A.R. Abate, and D.J. Durian}
\affiliation{UCLA Department of Physics \& Astronomy, Los Angeles, CA
90095-1547}


\date{\today}

\begin{abstract}
     The dynamics of a sphere fluidized in a nearly-levitating upflow
     of air were previously found to be identical to those of a
     Brownian particle in a two-dimensional harmonic trap, consistent
     with a Langevin equation [Ojha {\it et al.}, Nature {\bf 427},
     521 (2004)].  The random forcing, the drag, and the trapping
     potential represent different aspects of the interaction of the
     sphere with the air flow.  In this paper we vary the experimental
     conditions for a single sphere, and report on how the force terms
     in the Langevin equation scale with air flow speed, sphere
     radius, sphere density, and system size.  We also report on the
     effective interaction potential between two spheres in an upflow
     of air.
\end{abstract}




\maketitle



\section{Introduction}

One of the great challenges in physics today is to understand the
dynamics of driven nonequilibrium systems~\cite{Ruelle}.  This is
particularly important in soft-matter physics, because the materials
often have a delicate mesoscopic structure that is easily perturbed
far from equilibrium.  There, an understanding of the microscopic
dynamics is crucial for a fundamental understanding of macroscopic
behavior.  Granular materials are an excellent example of this
point~\cite{JNB,Duran}.  When subjected to strong driving forces,
granular systems exhibit gas- or liquid-like behavior at the
macroscopic scale and strong velocity fluctuations and collisions at
the grain-scale.  The microscopic fluctuations are created by the act
of flowing, and, at the same time, are responsible for the dissipation
that limits the rate of flow.  The difficulty of treating the
fluctuations is one reason why granular mechanics remains a forefront
research topic, and why engineering systems are alarmingly prone to
failure.

One way to characterize the microscopic dynamics in a granular gas or
liquid is by the distribution of speed fluctuations.  This has a long
history, and is associated with attempts to develop a system of
partial differential equations describing granular
hydrodynamics~\cite{Bagnold,Ogawa,JenkinsSavage,Haff,Reydellet,Bocquet}.
  The average kinetic energy associated with speed fluctuations has
come to be known as the ``granular temperature'', in loose analogy
with kinetic theory of gases.  An interesting line of research has
been to explore the extent to which this analogy holds, {\it i.e.}
the extent to which statistical mechanical concepts for true
thermal systems can be used to describe granular fluctuations.
For dilute or two-dimensional systems it is relatively
straightforward to track grain motion by video techniques.
Experimentalists have thus studied whether or not speed
distributions are Gaussian, and whether or not equipartion is
obeyed~\cite{Pouligny,Ippolito,Olafsen,Feitosa,Baxter}. Recently
we did the same for a very dilute system, consisting of only a
single grain, driven by a steady upflow of gas~\cite{Rajesh}. Part
of our motivation was to isolate the role of gas-mediated
interactions from collisional and cohesive interactions in bulk
gas-fluidized beds, which is a topic of long-standing
importance~\cite{Geldart}.  By measuring the time-dependent
dynamics, as well as the usual speed distribution, and by
performing auxiliary mechanical measurements, we were able to
demonstrate that the motion of the sphere is identical to that of
a Brownian particle in a two-dimensional harmonic trap.  For such
a system the thermal analogy is perfect.

In this paper we exploit the thermal analogy to deduce quantitative
information about interactions in gas-fluidized systems.  Now that the
tools of statistical mechanics are at our disposal, we may deduce the
salient features of the forces acting on a sphere from measurements of
position and speed statistics.  Besides providing additional data and
a more detailed description than in Ref.~\cite{Rajesh}, this follows
through on our original motivation to study the fundamental forces at
play in gas-fluidized beds.  Our statistical mechanical approach is
completely orthogonal to traditional wind-tunnel
measurements~\cite{Leweke}, and provides a clean decomposition of
gas-mediated interactions into three distinct contributions.  We begin
with a discussion of statistical mechanics and the Langevin equation
of motion, both to review prior findings and to define notation for
use here.  After describing our experimental apparatus, we then
present data pertaining to the effective temperature and its scaling
with system parameters, all for a one-sphere system.  Lastly, we turn to
interactions of a sphere with both the container boundary as well as
with a second sphere.

\section{Langevin Equation\label{langevinequation}}

The particles of interest are spheres of mass $m$, diameter
$D=2R_{d}$, and moment of inertia $I$.  They roll without sliding, so
their kinetic energy is $K={1\over2}(m+I/{R_{d}}^{2})v^{2}$.  In order
to characterize the motion entirely in terms of position, velocity,
and acceleration vectors, $\{{\bf r}(t),{\bf v}(t),{\bf a}(t)\}$, we
define an effective inertial mass and density as
$m_{e}=m+I/{R_{d}}^{2}$ and $\rho_{e}=m_{e}/[(4\pi/3){R_{d}}^{3}]$,
respectively.  As shown in Ref.~\cite{Rajesh}, the equation of
translational motion of the rolling gas-fluidized sphere is
\begin{equation}
     m_{e}{\bf a}(t) = -{\bf\nabla}V({\bf r}) -
     m_{e}\int_{-\infty}^{t}\Gamma(t-t'){\bf v}(t')dt' +
     {\bf F}_{r}(t).
\label{langevin}
\end{equation}
This is recognized as Newton's Second Law, where the right hand side
is the sum of forces acting on the sphere.  The first term is the
gradient of an effective potential; for a harmonic spring this force
is $-K{\bf r}(t)$.  The second term represents the drag force, where
$\Gamma(t)$ is the memory kernal.  In Ref.~\cite{Rajesh} it was shown
to be exponential,
\begin{equation}
     \Gamma(t)=\Gamma_{\circ}\gamma_{\circ}\exp(-\gamma_{\circ}t).
\label{kernal}
\end{equation}
Thus $1/\gamma_{\circ}$ is a time scale representing the duration of
the memory; $1/\Gamma_{\circ}$ is a time scale such that the drag
force has a typical value of $-m_{e}\Gamma_{\circ}{\bf v}$.  The final
term in Eq.~(\ref{langevin}) is a time-varying random force ${\bf
F}_{r}(t)$.  As shown in Ref.~\cite{Rajesh}, the components of ${\bf
F}_{r}(t)$ have Gaussian distributions and exponential temporal
autocorrelations.  In particular, it was demonstrated that the random
and drag forces are related according to the Fluctuation-Dissipation
Relation (FDR)~\cite{kubo}:
\begin{equation}
     \langle {\bf F}_{r}(t')\cdot{\bf F}_{r}(t)\rangle  =
     2 m_{e} kT \Gamma(t-t'),
\label{FDR}
\end{equation}
where $kT=m_{e}\langle v^{2}\rangle/2$ is the effective (granular)
temperature.  Note that the two momentum degrees of freedom each have
$kT/2$ of energy, consistent with the equipartition theorem.
Satisfaction of the FDR means that the particle dynamics are identical
to those of a thermal Brownian particle; therefore,
Eq.~(\ref{langevin}) is truly the Langevin Equation.  Even though the
rolling sphere is a driven far-from-equilibrium system, statistical
mechanics holds unchanged except that the value of the effective
temperature is not the thermodynamic temperature of the apparatus.

\section{Experimental Details}

Our methods are identical to those first reported in
Ref.~\cite{Rajesh}.  The fluidization apparatus is built around a
12-inch diameter brass sieve, with 300~$\mu$m wire mesh spacing
and with 4~inch high side wall.  The full sieve is usually used,
but occasionally a cylindrical insert is placed concentrically in
order to vary the radius $R_{cell}$ and/or the wall height.  The
wire mesh is flat and level, and is very fine compared to the
sphere size.  The sieve is mounted atop a 20~inch $\times$ 20~inch
$\times$ 4~foot tall windbox consisting of two nearly cubical
chambers separated by a perforated metal sheet.  In some of the
runs, a 1/2-inch thick foam air filter is sandwiched between a
second perforated metal sheet. Air from a blower is introduced to
the lower chamber through a flexible cloth sleeve.  The flow rate
is controlled by a variac.  The geometry of the windbox is
designed to achieve a uniform upwards air flow across the whole
area of the sieve.  This is verified and monitored with a hotwire
anemometer.

The sphere position is measured from digital images acquired at a
rate of 120 frames per second.  The camera has a resolution of
$640\times480$ pixels, and is mounted about 1 meter directly above
the sieve via a scaffolding attached to the windbox.  Two 18-inch
fluorescent lights are mounted just below the camera, such that
the illumination is uniform and the thresholded image of the
sphere appears as a white disk in a black background. In order to
achieve very long run times, using an ordinary personal computer,
we developed custom video compression and particle tracking
algorithms that permit real-time analysis without the need for
writing prohibitively large data-sets to hard-drive.  At heart is
a run-length encoding scheme: for each row, it's enough to note
the starting pixel and the segment length.  Since black pixels
have zero intensity, the sphere location is then computed as the
center of brightness of the entire thresholded image.

The sphere velocity and acceleration are found by postprocessing
position vs time data.  Specifically, we fit a third-order polynomial
to data within a window of $\pm4$ points.  Gaussian weighting that is
nearly zero at the edges is used to ensure continuity of the
derivatives.  This process also refines the position measurement.
In the end we achieve a resolution of $\pm0.05$~mm, which corresponds
to about 0.1$\%$ of the sphere diameter and about 0.08 pixels.

The specific spheres studies are listed in Table~\ref{balls}.  For
each, the allowed air speeds $u$ are bounded by 200 and 500~cm/s
depending on the sphere.  The range is limited because at lower
air speeds, the sphere occasionally rolls along its seem or along
the weave of the wire mesh.  At higher air speeds, the sphere
occasionally scoots or loses contact with the sieve.  In all
cases, the air speed is less than the terminal falling speed of
the sphere.  The Reynolds number based on sphere size is of order
$10^{4}$.  Thus the sphere sheds turbulent wakes, and this gives
rise to the stochastic motion.

\begin{table}[ht]
\begin{ruledtabular}
\begin{tabular}{lcc}
Sphere &  $\rho_{e}$ (g/cm$^{3}$) & D (cm)  \\ \colrule
king-pong & 0.122 & 4.41  \\
ping-pong & 0.146 & 3.80  \\
wood & 0.987 & 1.27$-$3.70 \\
polypropylene & 1.14 & 0.56$-$2.54 \\
nylon & 1.56 & 0.63$-$2.54 \\
\end{tabular}
\end{ruledtabular}
\caption{Inertial mass density and diameter for the various spheres.
The ping-pong and king-pong balls are both hollow plastic spheres,
with a 0.4~mm shell thickness; all others are solid.}
\label{balls}
\end{table}

\section{Effective Temperature\label{Teff}}

\begin{figure}
\includegraphics[width=3.00in]{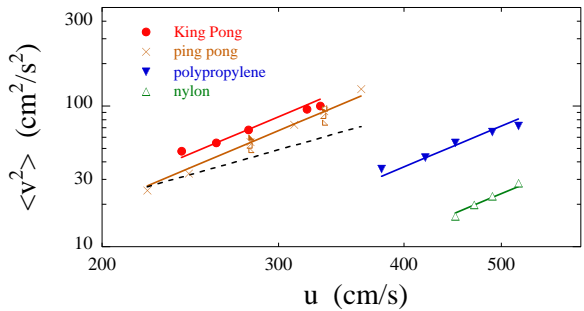}
\caption{The mean-squared speed of a rolling sphere vs the speed
of the upflow of air, for several types of sphere as labelled. The
solid lines are a best fit to cubic behavior, $\langle
v^{2}\rangle \sim u^{3}$.  The data are not consistent with the
dimensionally simpler scaling $\langle v^{2}\rangle \sim u^{2}$,
shown as a dashed line.  All data are for the full 12-inch sieve,
except the small right triangles for ping-pong balls in cells of
smaller radii.} \label{speeds}
\end{figure}

In this section we begin reporting on how the various terms in the
Langevin Equation scale with system parameters.  The first is the
effective temperature, given by the mean-squared speed as $kT =
m_{e}\langle v^{2}\rangle/2$.  Data for the mean-squared sphere speed is
shown as a function of air speed $u$ in Fig.~\ref{speeds} for various
types of sphere.  In all cases, the data are inconsistent with the
simplest dimensionally correct scaling, $\langle v^{2}\rangle \sim
u^{2}$.  Rather, the mean-squared speed appears to scale as the cube
of the air speed.  Thus $u^{3}/\langle v^{2}\rangle$ has units of speed
and presumably depends on physical characteristics of the sphere, the
fluidizing air, and gravity.

\begin{figure}
\includegraphics[width=3.00in]{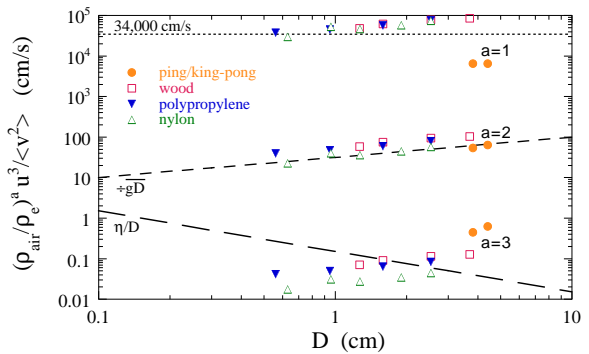}
\caption{Scaling of the characteristics speeds with the sphere
diameter, $D$, for several different spheres as labelled.  The
mean-squared speed $\langle v^{2}\rangle$ of the sphere is
proportional to the cube of the air speed, $u^{3}$; therefore, the
ratio of these quantities is a characteristic speed that reflects
both the sphere density and the dissipation mechanism.  The data
collapse is best when $u^{3}/\langle v^{2}\rangle$ is multiplied
by the square of the density ratio.  Then the value and form of
the characteristic speed are both consistent with $\sqrt{gD}$,
indicating that rolling friction is the dominant dissipation
mechanism.} \label{diam}
\end{figure}

To uncover the full scaling we first proceed by dimensional
analysis. Assuming that $\langle v^{2}\rangle$ decreases with
increasing sphere density, the combination
$(\rho_{air}/\rho_{e})^{a}u^{3}/\langle v^{2}\rangle$ is the
important characteristic speed, where the exponent $a$ of the
density ratio is to be determined.  We can conceive of only three
possibilities for the origin of this characteristic speed: the
speed of sound, 34,000~cm/s; a speed set by gravity and the sphere
size, $\sqrt{gD}$; and a speed set by air viscosity and sphere
size, $\eta/D$. To investigate, we compare these possibilities
with data for the characteristic speed vs sphere size in
Fig.~\ref{diam}, for several integer values of $a$.  We find that
the best data collapse is attained for $a=2$.  For that case the
value and functional form of the characteristic speed are both
consistent with $\sqrt{gD}$. Adjusting the numerical prefactor to
best match all the data, we thus find that the mean-squared speed
of a sphere is given by
\begin{equation}
     \langle v^{2}\rangle = 0.7\left({\rho_{air}\over\rho_{e}}\right)^{2}
     {u^{3} \over \sqrt{gD}}.
\label{v2}
\end{equation}

This observed scaling of the mean-squared sphere speed is
consistent with a simple model of the stochastic motion of the
sphere being driven by turbulence in the air.  The idea is to
balance the rate $P_{in}$ at which kinetic energy is transferred
from the air to the sphere with the rate $P_{out}$ at which energy
is dissipated by drag. Ignoring numerical factors, the latter is
the characteristic drag force times the characteristic sphere
speed: $P_{out}=(\rho_{e}D^{3}\Gamma_{\circ}v)v$, using the
notation of Section~\ref{langevinequation}.  The former is
$P_{in}=(\rho_{e}D^{3}\delta v^{2})\gamma_{\circ}$, where the term
in parenthesis is the kinetic energy change due to the shedding of
a wake and $\gamma_{\circ}=u/D$ is the rate at which wakes are
shed.  Assuming that the wake size scales with sphere size,
momentum conservation gives $\rho_{e}\delta v=\rho_{air}u$.  The
numerical prefactor is nontrivial, since it must depend on the
ratio of wake to sphere size and also on the fraction of momentum
in the plane of the sieve, transverse to the average air flow
direction.  Combining all these elements, the balance of power
input with power output is ${\rho_{air}}^{2}u^{3}/D =
{\rho_{e}}^{2}\langle v^{2}\rangle\Gamma_{\circ}$.  This is
identical to our data on the mean-squared speed, Eq.~(\ref{v2}),
provided that the drag amplitude scales as
$\Gamma_{\circ}\propto\sqrt{g/D}$ and that the memory decay rate
scales as $\gamma_{\circ}\propto u/D$.  Next we demonstrate that
these provisos both hold true.

\section{Drag and Random Forces}

\begin{figure}
\includegraphics[width=3.00in]{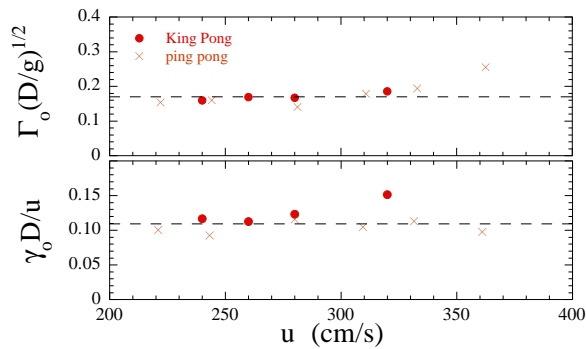}
\caption{Amplitude $\Gamma_{\circ}$ and decay rate
$\gamma_{\circ}$ of the memory kernal,
$\Gamma(t)=\Gamma_{\circ}\gamma_{\circ}\exp(-\gamma_{\circ}t)$, as
a function of air flow speed, for two different spheres as
labelled; these quantities are rendered dimensionless by
appropriate factors of sphere diameter, gravitational
acceleration, and air flow speed according to expectation.  The
dashed lines represent average values, 0.17 in the top plot and
0.11 in the bottom plot.} \label{gammas}
\end{figure}

Recall from Eqs.~(\ref{langevin}-\ref{FDR}) in
Section~\ref{langevinequation} that both the drag and random forces
are specified by an exponential memory kernal,
$\Gamma(t)=\Gamma_{\circ}\gamma_{\circ}\exp(-\gamma_{\circ}t)$.  In
Ref.~\cite{Rajesh} we found consistent values for $\Gamma_{\circ}$ and
$\gamma_{\circ}$ from two different methods.  The first was from the
velocity autocorrelation function using the Langevin equation.  The
second was from the amplitude and phase of the average response to a
small sinusoidal rocking of the entire apparatus at various
frequencies.  Here we employ the former method for both the ping-pong
and king-pong balls, as a function of air speed.  The results are
shown in Fig.~\ref{gammas}, made dimensionless according to the
expectations of Section~\ref{Teff}.  Specifically, the top plot
demonstrates that the drag amplitude behaves as expected:
\begin{equation}
     \Gamma_{\circ}=0.17\sqrt{g/D}.
\label{dragamp}
\end{equation}
The importance of $g$ suggests that rolling friction is the dominant
source of drag, as opposed to shear or compression of the air.
Perhaps we may identify the numerical prefactor of $g$ as a
coefficient of friction, $\Gamma_{\circ}=\sqrt{\mu g/D}$ with
$\mu=0.03$.  It would be interesting to investigate how $\mu$ changes
with mesh size and ball roughness.  The bottom plot of
Fig.~\ref{gammas} demonstrates that the memory decay rate also behaves
as expected:
\begin{equation}
     \gamma_{\circ} = 0.11 u/D.
\label{wake}
\end{equation}
This is consistent with earlier visualization and pressure
fluctuation studies, which found that the vortex shedding
frequency is $0.15u/D$ for Reynolds number in the range
$10^{3}-10^{6}$~\cite{achenbach,suryanarayana}.  Here,
Eq.~(\ref{wake}) means that a new wake is shed every time the air
flows a distance of about nine sphere diameters; equivalently, the
Strouhal number is ${\rm St}\equiv\gamma_{\circ}D/u=0.11$.

We emphasize that while the results of Eqs.~(\ref{dragamp}-\ref{wake})
directly specify the drag force, they also specify the random force
via the Fluctuation-Dissipation Relation Eq.~(\ref{FDR}).
The random driving and the drag forces are different
aspects of the same physical interaction between the sphere and the
turbulence it generates in the air.  To recap, the random force has
Gaussian components and an exponential temporal autocorrelation,
\begin{equation}
     \langle {\bf F}_{r}(t')\cdot{\bf F}_{r}(t)\rangle  =
     2 m_{e} kT \Gamma_{\circ}\gamma_{\circ}\exp[\gamma_{\circ}(t-t')],
\label{FDR2}
\end{equation}
where $kT=m_{e}\langle v^{2}\rangle/2$ is specified by Eq.~(\ref{v2}).

\section{Ball-Wall Interaction}

The potential $V(r)$ is the only part of the Langevin Equation not yet
discussed.  This can be deduced from the radial position probability
function, $P(r)$, using principles of statistical mechanics.  Namely,
the probability to find the sphere in a thin ring of radius $r$ is
proportional to the ring radius times a Boltzmann factor,
\begin{equation}
     P(r) \propto r \exp[-V(r)/kT],
\label{V}
\end{equation}
where $kT$ is the effective temperature discussed in
Section~\ref{Teff}.  In Ref.~\cite{Rajesh}, the sphere was found
most frequently near the center of the cell such that the $x$ and
$y$ distributions were nearly Gaussian and $P(r)\approx
(2r/\langle r^{2}\rangle)\exp(-r^{2}/\langle r^{2}\rangle)$.  This
means that the interaction potential is nearly harmonic,
$V(r)\approx Kr^{2}/2$.  The value of $\langle r^{2}\rangle$ gave
a spring constant that was verified by an auxiliary mechanical
tilting measurement.  Here, we examine the shape of the potential
more closely, and we explore its behavior as a function of system
parameters.

\begin{figure}
\includegraphics[width=3.00in]{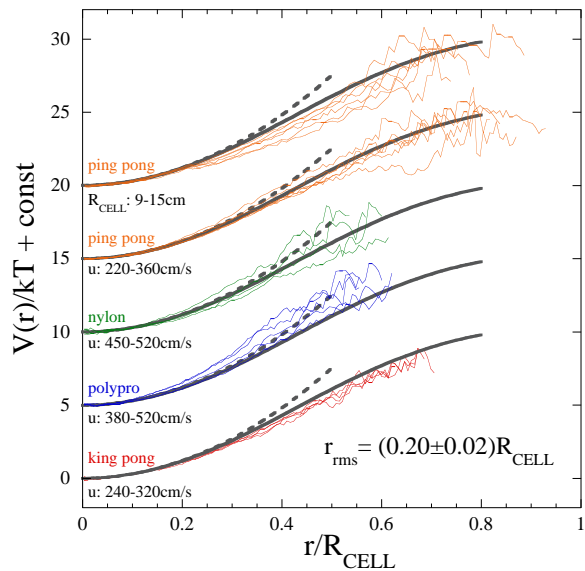}
\caption{Interaction potential between a variety of spheres and
the walls of the container, vs distance from the center of the
cell scaled by cell radius.  The top data set is taken at constant
air flow speed, while all others are taken at constant (full) cell
size. The dashed curves represent a harmonic potential,
$V_{h}(r)/kT=30(r/R_{cell})^{2}$.  The solid curves represent an
empirical average of all the data,
$V(r)/kT=30(r/R_{cell})^{2}/[1+2(r/R_{cell})^{3}]$.  An arbitrary
constant offset was added in order to separate the different
datasets.} \label{ballwall}
\end{figure}

Radial position probability data for all runs are converted to the
interaction potential via Eq.~(\ref{V}), and displayed altogether in
Fig.~\ref{ballwall}.  The potential is left in units of $kT$, and the
radial position is scaled by the cell radius, $R_{cell}$, for
clarity.  Remarkably, the potential is given by the same empirical
form independent of sphere size, cell radius, and air flow speed:
\begin{equation}
     V(r)/kT = {30(r/R_{cell})^{2} \over 1+(r/R_{cell})^{3}}.
\label{Vemp}
\end{equation}
In particular, the rms radial position of a sphere is always set by
the cell size, $\sqrt{\langle r^{2}\rangle}=(0.20\pm0.01)R_{cell}$.
The harmonic form of the potential also softens away from the center.
It actually becomes attractive very close to the walls, strong enough
to occasionally trap an unwary sphere that wanders too far from home.

The geometric scaling of the potential with cell size, independent of
air flow speed, leads us to believe that the origin of the behavior
lies in the interaction of the shed vortices with the boundary of the
cell.  This is bolstered by other observations as well.  First, even
very slight imperfections in the circularity of the cell can break the
radial symmetry of the position distributions.  Second, placing a hand
or other object downstream from the sphere affects its position
distribution as well.  Evidently, the vortex street is connected to
the sphere such that force can be exerted on the sphere via
perturbation to the vortices.

One possible picture for how the vortex street senses the wall is that
the transverse extent of the vortices grows linearly with distance
downstream.  Then the sphere could sense its position from the height at
which the expanding vortices hit the boundary.  See Figs.~31, 55, 56,
172, and 173 of Ref.~\cite{vandyke} for photographs of the vortices
behind various objects at approximately the same Reynolds number as
here.  Another possible picture is that the background flow, while
homogeneous near the sieve, develops large-scale structure downstream
that grows from the edge inwards.  Then the sphere could senses its
position from from the height at which its vortices merge with the
dome of turbulent structure above.  See Fig.~152-153 of
Ref.~\cite{vandyke} for photographs of the isotropic turbulence behind
a grid and it's evolution downstream.

We performed a few tests in attempt to clarify the physical pictures.
First we increased and decreased the wall height to considerable
extent.  This had no influence on the sphere position statistics,
which seems to rule out the growing-vortex scenario.  Our second test
was to stretch a fine netting across the top of the sieve.  We hoped
that this would affect the rate of vortex shedding or the way the
vortex street is connected back to the sphere.  However, it had no
influence on the sphere position statistics either.  Thus, we must
leave the origin of the geometric nature of the sphere-wall
interaction potential as something of a mystery.  Flow visualization
may be helpful.  We close by emphasizing that, whatever its origin,
the sphere is repelled by the cell wall in a way that, remarkably, can
be described by a potential energy and a corresponding conservative
force.

\section{Ball-Ball Interaction}

In the remainder of this paper we report on the air-mediated
interaction between two spheres rolling in the same nearly-levitating
upflow of air.  Throughout, we set the air flow to 280~cm/s, before
adding spheres.  As above, we shall see that this may be studied using
position probability data and statistical mechanics.  And just as for
the ball-wall force, we shall see that the ball-ball force is
repulsive.  Naively one might expect a Bernoulli-like attraction, just
as when air is blown between two objects.  However it's immediately
obvious from visual inspection that here the two spheres repel.  Only
rarely do they collide, with physical contact between their surfaces;
they never stick; they accelerate apart after close approach.

\begin{figure}
\includegraphics[width=3.00in]{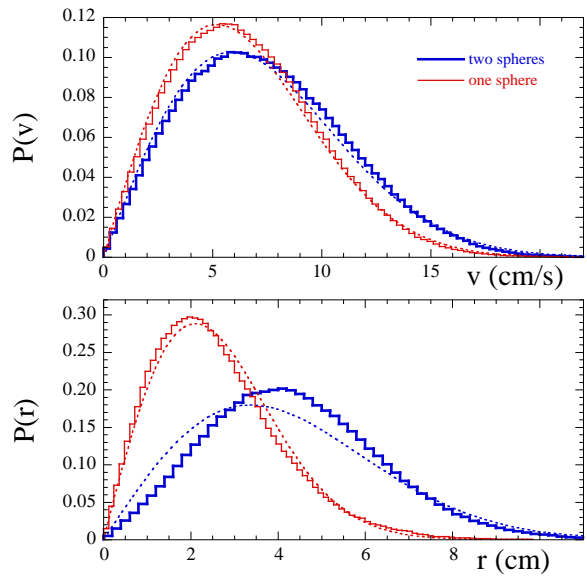}
\caption{Speed and radial position probability functions for
one and two spheres rolling in the same upflow of air.  Note that the
speed distributions are Gaussian in both cases, as shown by the dashed
curves.  By contrast, the radial position function becomes non-Gaussian
when a second sphere is added to the system.}
\label{vrdist}
\end{figure}

To begin we display speed and radial position probabilities in
Fig.~\ref{vrdist}.  The light curves are for a single ball in the same
air flow, for comparison.  As above and in Ref.~\cite{Rajesh}, the $x$
and $y$ components of velocity and position are all Gaussian.  When a
second sphere is added, we verify that the velocity and position
distributions remain radially symmetric and identical for each sphere.
The top plot of Fig.~\ref{vrdist} demonstrates that the average speed
distribution of the two spheres remains Gaussian.  Thus the
mean-squared speed can be used to define an effective temperature, as
before for a single sphere.  However, this temperature increases when
a second sphere is added, even though the flux of air remains
unchanged.  Evidently Eq.~(\ref{v2}) holds in detail only for a
one-sphere system.  The reason may be that, due to a decrease in free
area, the air flow speed around the two spheres is greater than when
only one is present.  It may also be that the process of energy
injection via vortex shedding is altered.  The bottom plot in
Fig.~\ref{vrdist} demonstrates that the radial position probability
becomes non-Gaussian when a second sphere is added.  Each sphere
spends less time in the very center of the cell, due to mutual
repulsion, with the rms radial position increasing from 2.8~cm to
4.8~cm when a second sphere is added.  As before, the spheres still
are repelled from the cell wall as though in a harmonic trap.

\begin{figure}
\includegraphics[width=3.00in]{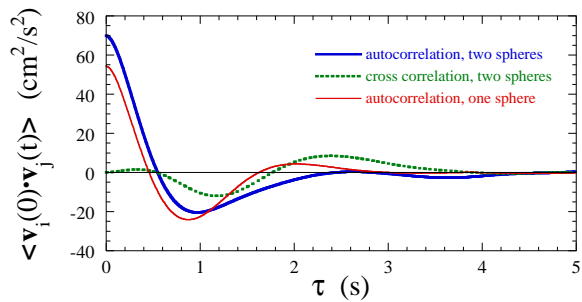}
\caption{Velocity correlations between spheres $i$ and $j$, for
one- and two-sphere systems as labelled.  Note that the
crosscorrelation vanishes at $\tau=0$, which is required if
statistical mechanics is to be invoked.} \label{corr}
\end{figure}

For statistical mechanics to be useful for studying the sphere-sphere
repulsion, it is required that the velocity components be Gaussian as
demonstrated above.  It is also required that there be no correlation
between the instantaneous velocities of the two spheres.  To check
this, we compute temporal velocity correlation functions and plot the
results in Fig.~\ref{corr}.  The velocity autocorrelation for a single
sphere alone in the cell is shown by a light curve, for comparison.
The velocity autocorrelation for each sphere, when two are present, is
shown by a heavier curve.  It decays over the same time scale as the
one-sphere autocorrelation, though the oscillations are less pronounced.
The cross correlation between the velocities of the two spheres is
shown by a dashed curve.  It too oscillates and decays over the same
time scale as the autocorrelations.  But, crucially for us, it
vanishes at $\tau=0$.  Thus the instantaneous equal-time velocities of
the two spheres are indeed uncorrelated as required.

With the above preliminaries established, we may now exploit the
principles of statistical mechanics in order to deduce the
sphere-sphere interaction potential $V_{ss}(\rho)$, where $\rho$
is the distance between the centers of the two spheres.  The idea
is to compute the sphere-sphere separation probability in terms of
both the overall harmonic confining potential and the unknown
$V_{ss}(\rho)$.  This is accomplished by summing the Boltzmann
factors for all the ways of arranging the spheres with the desired
separation:
\begin{eqnarray}
     P(\rho) & \propto & \int{\rm d}x{\rm d}y{\rm d}\theta
                 \exp\bigg[-{1\over2}K\bigg(x^{2}+y^{2} + \\
     &  & (x+\rho\cos\theta)^{2}+(y+\rho\sin\theta)^{2}\bigg)/kT\bigg]
\nonumber \\
     &  & \times \exp[-V_{ss}(\rho)/kT] \nonumber \\
     & \propto & \exp\left[-\left( {1\over4}K\rho^{2}+V_{ss}(\rho)
     \right)/kT\right].
\label{Prho}
\end{eqnarray}
One may differentiate this expression to show that the peak in
$P(\rho)$ is where $-{\rm d}V_{ss}/{\rm d}\rho = K\rho/2$, which is a
statement of force balance when each sphere is $\rho/2$ from the
center of the cell.  Since the spring constant $K$ is known from the
one-sphere experiment, and since the temperature $kT$ is known from the
mean-squared speeds, the functions $P(\rho)$ and $V_{ss}(\rho)$ may be
deduced one from the other.

\begin{figure}
\includegraphics[width=3.00in]{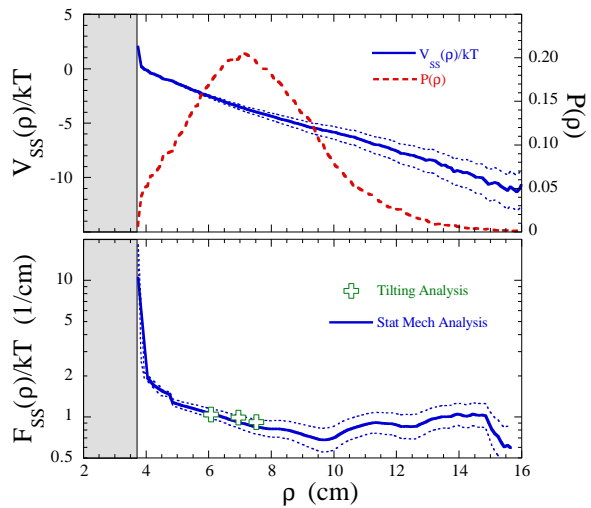}
\caption{Interaction between two spheres as a function of
their separation $\rho$.  Using Eq.~(\protect{\ref{Prho}}), the
potential $V_{ss}(\rho)$ is inferred from the separation probability
$P(\rho)$; these functions are both shown in the upper plot.
Systematic uncertainty in $V_{ss}(\rho)$ is indicated by dotted curves; it
is due to statistical uncertainty in the value of $K$.  The repulsive
force is shown in the bottom plot, as obtained both from $V_{ss}(\rho)$ and
from an auxilliary mechanical measurement.}
\label{ballball}
\end{figure}

The separation probability $P(\rho)$ is readily found from the video
data for the position of each sphere vs time.  Results are
displayed by a dashed curve on the right axis of the upper plot in
Fig.~\ref{ballball}.  The probability rises abruptly from zero at a
separation equal to the sphere diameter.  It reaches a peak near
$\rho=7$~cm, and then gradually decays again toward zero.  The
sphere-sphere potential $V_{ss}(\rho)$ can then be obtained from $P(\rho)$
using Eq.~(\ref{Prho}).  Results are shown by a solid curve on the
left axis of the upper plot in Fig.~\ref{ballball}.  The precipitous
drop of $V_{ss}(\rho)$ near contact indicates a hardcore repulsion.  The
more gradual drop at larger separations indicates a softer repulsion.

The actual force of repulsion may be found by differentiating,
$F_{ss}(\rho)=-{\rm d}V_{ss}/{\rm d}\rho$.  Results are shown by
the solid curve in the lower plot of Fig.~\ref{ballball}.  There
is a hardcore repulsion, followed by a nearly constant-force
repulsion when the sphere centers are separated by more than two
diameters. Expressing the interaction in terms of a force allows
us to perform a check using an auxiliary mechanical measurement of
the response to tilting the entire apparatus by a fixed angle
$\theta$ away from horizontal.  This causes a constant component
of gravity, $mg\sin\theta$, within the plane and breaks the radial
symmetry; note that here $m$ is the true mass, not the effective
inertial mass.  Then we measure the probability $\phi(x,y)$ for
finding a sphere at a given position, where the origin of the
coordinate system is at the center of the cell and where gravity
acts in the $+\hat y$ direction.  This probability has two peaks,
at coordinates $(\pm\rho/2,y_{p})$, separated by distance $\rho$.
Assuming only that the wall repulsion acts in the radial
direction, the statement of force balance at the peaks of
$\phi(x,y)$ gives the sphere-sphere repulsive force as
\begin{equation}
     F_{ss}(\rho)=\left({1\over2}\rho/y_{p}\right)mg\sin\theta.
\label{F}
\end{equation}
In practice, to achieve a wider range in separations, we tilt the
apparatus by 0.013~rad and use cells of three different diameters:
20, 25, and 30~cm.  Observations then give the repulsive force at
three different separations as shown in the lower plot of
Fig.~\ref{ballball}.  Evidently the agreement with the results from
statistical mechanics is very good.  This gives confidence in the use
of statistical mechanics to deduce the full form of the repulsive
sphere-sphere interaction.

\section{Conclusion}

We have exploited the thermal-like behavior of a single gas-fluidized
sphere to deduce the nature of the forces dictating its motion.  All
these forces are mediated by turbulence in the gas, but can be
decomposed into distinct contributions.  Due to randomness in the
shedding of turbulent wakes, there is a rapidly varying random force
specified by Eqs.(\ref{dragamp}-\ref{FDR2}).  By virtue of the
Fluctuation-Dissipation Relation, and
Eqs.(\ref{langevin}-\ref{kernal}), these results also fully specify a
velocity-dependent drag force that damps rolling motion.  The apparent
interaction of the wakes with the cell boundary gives rise to a
nearly-harmonic force that keeps the rms sphere position at about one
fifth the cell radius, no matter how the system parameters are
changed.  The effective temperature, set by the mean-squared speed in
Eq.~(\ref{v2}), is a key parameter in these forces.  When a second
sphere is added, the thermal analogy still holds and these forces
change only in detail.  In addition, there is a gas-mediated repulsion
acting between the spheres that is nearly constant beyond a few
diameters of separation and that grows stronger near contact.

We thank L. Bocquet, R.F. Bruinsma, P.G. de~Gennes, D. Frenkel, J.B.
Freund, D. Levine, and M.A. Rutgers for useful conversations.  We
thank P.K. Dixon and P.-A. Lemieux for assistance with digital video
techniques, and A.J. Liu for assistance with the Langevin analysis.
This work was supported by NSF through grant numbers DMR-9623567 and
DMR-0305106.

\bibliography{AirBallRefs}

\end{document}